\documentclass[11pt]{article}
\nonstopmode

\usepackage{natbib,hypernat}
\usepackage{graphicx,wrapfig,graphics} 
\usepackage[nodayofweek]{datetime}
\usepackage{array}
\usepackage[utf8]{inputenc}
\usepackage{hyperref,url}
\usepackage{parskip}
\usepackage{listings}
\usepackage{verbatim}

\usepackage{tikz}
\usetikzlibrary{positioning}
\usetikzlibrary{calc}
\usetikzlibrary{chains}
\usetikzlibrary{shapes.misc}
\usetikzlibrary{shapes.symbols}
\usetikzlibrary{shapes.geometric}
\usetikzlibrary{fit}
\usetikzlibrary{shadows}

\usepackage[subfigure]{ccaption}
\newfloatlist{proglisting}{tpb}{listings}{Listing}
\newsubfloat{proglisting}

\pgfdeclarelayer{background}
\pgfdeclarelayer{foreground}
\pgfsetlayers{background,main,foreground}

\pgfrealjobname{main}

\usepackage{hyperref}
\usepackage{subfigure}

\title{GPU~Scripting and Code~Generation with PyCUDA}

\author{
Andreas Klöckner\footnote{%
  Courant Institute of Mathematical Sciences,
  New York University,
  New York, NY 10012}\,\
\and
Nicolas Pinto\footnote{%
  Brain and Computer Sciences,
  Massachusetts Institute of Technology,
  Cambridge, MA 02139}\,\
\and
Bryan Catanzaro\footnote{%
  Electrical Engineering and Computer Sciences,
  University of California,
  Berkeley, CA 94720}\,\
\and
Yunsup Lee \addtocounter{footnote}{-1}\footnotemark{}%
\and
Paul Ivanov\footnote{%
  Redwood Center for Theoretical Neuroscience,
  University of California,
  Berkeley, CA 94720}\,\
\and
Ahmed Fasih\footnote{%
  Department of Electrical and Computer Engineering,
  Ohio State University,
  Columbus, OH 43210}\,\
}

\definecolor{darkgreen}{rgb}{0,0.4,0}

\newcommand{\ednote}[1]{}
\newcommand{\authnote}[1]{}

\usepackage{amsmath}

\colorlet{codeback}{gray!20}

\begin{document}
\maketitle

%
%
%
%



\ednote{A paragraph with no section heading to introduce the chapter.}

High-level scripting languages are in many ways polar opposites to
GPUs. GPUs are highly parallel, subject to hardware subtleties, and
designed for maximum throughput, and they offer a tremendous advance
in the performance achievable for a significant number of
computational problems. On the other hand, scripting languages such as
Python favor ease of use over computational speed and do not generally
emphasize parallelism. PyCUDA is a package that attempts to join the
two together. This chapter argues that in doing so, a programming
environment is created that is greater than just the sum of its two
parts.

We would like to note that nearly all of this chapter applies in
unmodified form to PyOpenCL, a sister project of PyCUDA, whose goal it
is to realize the same concepts as PyCUDA for OpenCL.

\section{Introduction, Problem Statement, and Context} \ednote{This
section should include a description of the general problem being
solved and how this technology helps to improve the state of the art,
but do so at two levels, one targeting nondomain experts, and others
for readers in this field.}

How are computational codes created? Their life cycle often begins
with a quick proof of concept in a high-level language such as MATLAB
or Python that aims to examine the suitability of the proposed method
for the application problem at hand. Once this is established, the
problem size is scaled up, and with it, the demands for execution
speed grow.  In principle, it is not desirable that the working
proof-of-concept code would need to be changed merely to squeeze out
better computational performance. Unfortunately, at present, very
few workloads can be scaled in this fashion, so eventually the need
arises to employ a language that produces efficient machine code.

The desire to make this transition as seamless as possible quite
naturally leads to hybrid code, in which only those pieces requiring
improved performance are changed, and the remainder is kept as-is.
These approaches are not new (cf. e.g. MATLAB's Mex or Python's F2Py).
Unfortunately, they are not as wide-spread as they could be because
they come with a significant complexity burden.  GPU computing adds a
new facet to this issue, as GPU programs are always hybrids,
even though NVIDIA's CUDA Runtime programming interface goes to
considerable lengths to paper over this fact.  GPU programs are thus
very naturally split between performance-hungry and
performance-indifferent parts, usually along the same lines as the
CPU-GPU boundary. This is fortunate, as one may easily substitute the
performance-indifferent part of the hybrid, taking advantage of the
already well-defined interface between the two. PyCUDA fits exactly
into this niche and allows codes written in this high-level language
to obtain GPU performance from within existing Python programs with a minimum
of effort.

But in addition to a practical route to high performance for existing
codes, PyCUDA also has much to offer to the seasoned GPU programmer
creating full-scale, production codes.  Many of these advanced
benefits are described in detail in a recent manuscript
\cite{kloeckner_pycuda_2009}, which focuses on the capability of
generating GPU code at run time and represents a more academic
discussion of the software engineering aspects of GPU programming and what
PyCUDA does to address them. This chapter on the other hand provides a
hands-on perspective of the things one can do with PyCUDA, and how.

\section{Core Method} \ednote{Concise overview of underlying methods
used in this solution.  One of the key objectives of this section
should be to provide enough information to nondomain experts to allow
them to make a determination whether they should read the rest of the
chapter.  In particular, readers not from your field will try to see if
the techniques in your chapter will be applicable to their problems.
It is therefore important to have a good illustration that allows
readers from other fields to quickly determine if the problem being
solved is similar to the ones they are trying to solve in their own
fields.}

\label{sec:core-method}


A common lament in the field of scientific computing concerns the
ever-widening gap between hypothetical machine capabilities and the
effort an individual programmer is able to spend to move a computational
application towards exploiting a machine to the full extent of its
capability. GPUs obviously influence this balance, by increasing the
maximum capability, by increasing the performance that is obtained
with an ``average'' amount of effort, and by requiring a different set
of skills to achieve genuinely high performance.

More sophisticated tools, compilers and libraries are generally hoped
to level this field, enabling users to achieve good results even with
modest investment. PyCUDA is a contribution to this discussion of
tools for GPU computing. PyCUDA (and likewise its sister project
PyOpenCL) has a two-fold aim. First, it aims to simplify the usage of
\emph{existing} basic concepts of CUDA C. Importantly, it does
\emph{not} attempt to change or reinvent the basic notions of GPU
programming, but instead, as a foundation for further tools, just
exposes them as-is through a carefully engineered interface. Key
features of this interface include generous use of sensible defaults
(all of which can be overridden), automatic error checking, and
resource management.  Second, and strictly \emph{on top} of the first,
basic layer, PyCUDA provides abstractions that support a number of
very common usage patterns involving a GPU equivalent of NumPy
\citep{oliphant_numpy_2006} arrays.

As a particular consequence of the design choice to leave as much of
the underlying concepts in place, device kernels occur in PyCUDA
programs and PyOpenCL as simple strings. This is practical because
Python allows so-called triple-quoted strings which may extend across
line boundaries, thus appearing as contiguous code blocks without
intervening string terminators or other traces of the host-level
language. A different approach is pursued by the Clyther 
project\footnote{\url{http://clyther.sourceforge.net/}} and similar efforts,
which let the user write kernels in the host language and then
translate them to the device language (C/C++) in some manner.  Each
approach represents one end of a trade-off. It is obviously appealing
to let the user program in one single language (the host language)
rather than forcing her to master two separate ones. PyCUDA and
PyOpenCL sacrifice this advantage in favor of passing on the concepts
of the underlying programming system in as unmodified a manner as
possible in order to provide direct applicability of existing CUDA/OpenCL
documentation and to decrease the maintenance burden as the underlying
system evolves. One additional factor entering the trade-off is the
ease with which kernel code can be automatically generated (see
Section \ref{sec:pycuda-code-generation}).

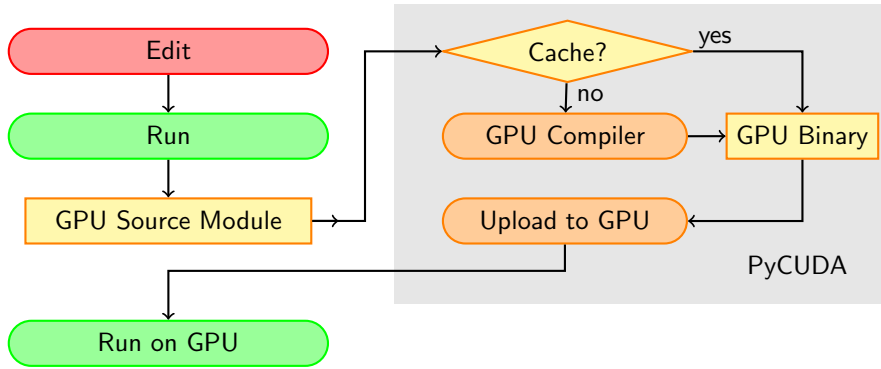
\begin{figure}[h]
  \centering
  \beginpgfgraphicnamed{pycuda-workflow}
  \begin{tikzpicture}[
    tbox/.style={minimum size=6mm,text height=1.5ex,text depth=0.25ex},
    stepbox/.style={rounded rectangle,thick,tbox,minimum width=4.5cm},
    sidestepbox/.style={rounded rectangle,thick,tbox,minimum width=3.5cm},
    objbox/.style={rectangle,thick,tbox},
    myarr/.style={thick},
    font=\sffamily\small,scale=0.7,
  ]
    \node[stepbox,draw=red,fill=red!40]
      (edit) {Edit} ;

    \draw [fill=gray!20,draw=none] (edit.east) +(1.25,0.9) rectangle +(10.5,-4.8)
      node [pos=0.95,anchor=south east] {PyCUDA};

    \node[stepbox,draw=green,fill=green!40,below=0.5 of edit]
      (run) {Run} ;

    \draw [->,myarr] (edit) -- (run) ;

    \node[objbox,draw=orange,fill=yellow!40,below=0.5 of run,minimum width=3.8cm]
      (smod) {GPU Source Module} ;
    \draw [->,myarr] (run) -- (smod) ;

    \draw [->,myarr] (smod.east) -- +(0.5,0) ;

    \node[diamond,aspect=4,draw=orange,fill=yellow!40,right=1.5 of edit,thick]
      (cache) {Cache?} ;
    \draw [->,myarr] (smod.east) -- +(1,0) |- (cache) ;

    \node[sidestepbox,draw=orange,fill=orange!40,right=1.5 of run]
      (nvcc) {GPU Compiler} ;
    \draw [->,myarr] (cache) -- (nvcc) node [pos=0.4,anchor=west] {no};

    \node[objbox,draw=orange,fill=yellow!40,right=0.5 of nvcc]
      (cubin) {GPU Binary} ;
    \draw [->,myarr] (nvcc) -- (cubin) ;

    \draw [->,myarr] (cache) -| (cubin) node [pos=0.1,above=-0.1cm] {yes} ;

    \node[sidestepbox,draw=orange,fill=orange!40,below=0.5 of nvcc]
      (upload) {Upload to GPU} ;
    \draw [->,myarr] (cubin) |- (upload) ;

    \node[stepbox,draw=green,fill=green!40,below=1 of smod]
      (gpurun) {Run on GPU} ;
    \draw [->,myarr] (upload.south) -- ++(0,-0.5) -| (gpurun) ;

  \end{tikzpicture}
  \endpgfgraphicnamed
  \caption{Workflow of PyCUDA GPU program compilation. PyCUDA aims to maintain
    a scripting-like ``edit-run-repeat'' style of working for the user. The
    compilation and caching operations in the gray box are performed without
    user involvement.}
  \label{fig:pycuda-workflow}
\end{figure}

Programming in scripting languages (like Python) tends to be quite
satisfying to the programmer because of near-immediate response during
development, the possibility of interactive exploration, and good
error reporting. Although PyCUDA gives the user access to a compiled
language (CUDA C), it attempts to avoid the development iteration
penalty commonly associated with compiled languages, instead retaining
the satisfaction and immediacy of scripting. First of all, it makes
invocations to the CUDA C compiler fast and transparent. To this end,
it employs a compiler caching mechanism that is pictured in Figure
\ref{fig:pycuda-workflow}. As it is likely that only one GPU kernel is
being changed in each development iteration, the user only needs to
wait for compilation of this one kernel. The loading of all other
kernels in the code will be near-instantaneous thanks to PyCUDA's
caching mechanism.

We have thus completed an initial description of PyCUDA's goals and
usage patterns. The next section will, by means of a number of
concrete examples of increasing complexity, show how common tasks can
be accomplished in PyCUDA. Section \ref{sec:eval} will then briefly
reflect on what was shown, and we close in Section \ref{sec:future}
with a few remarks and ideas for future work.

\section{Algorithms, Implementations, and Evaluations}
\ednote{This is the core of the chapter. This can be a multi-subsection section
It presents key insights from algorithm implementations and
evaluations along the way. Each subsection can give the algorithm, 
implementation, and measured benefit of a version of your solution. 
The section should be written to maximize the likelihood that someone 
can reproduce your success.}

In this section, we will take the reader on a brief journey through
different aspects of programming GPUs using PyCUDA, starting with a
basic hello-world example in Section \ref{sec:pycuda-hello-world},
through more advanced aspects in the following sections.

\subsection{The Basics of GPU programming with PyCUDA}
\label{sec:pycuda-hello-world}

Perhaps the simplest useful program that can be written using PyCUDA
is shown in Listing \ref{lst:pycuda-demo}, which we will discuss
here step-by-step.

\begin{proglisting}[t]
  \begin{minipage}{\textwidth}
  \tikz [remember picture] \coordinate (listing-top-left) ;
\begin{lstlisting}
import pycuda.driver as cuda
import pycuda.autoinit
import numpy

a = numpy.random.randn(4,4) \
        .astype(numpy.float32)
a_gpu = cuda.mem_alloc(a.nbytes)
cuda.memcpy_htod(a_gpu, a)

mod = cuda.SourceModule("""
  __global__ void multiply2(float *a)
  {
    int i = threadIdx.x + threadIdx.y*4;
    a[i] *= 2;
  }
  """)

func = mod.get_function("multiply2")
func(a_gpu, block=(4,4,1))

a_doubled = numpy.empty_like(a)
cuda.memcpy_dtoh(a_doubled, a_gpu)
print a_doubled
print a
\end{lstlisting}
  \begin{tikzpicture}[remember picture, overlay]
    \filldraw[rounded corners, opacity=0.3, color=red] 
    (listing-top-left) ++(0.2,-4.5) rectangle +(0.95\textwidth, -5.5em) 
    node [pos=0.98,anchor=south east,text=black,opacity=1,
    font=\small\sffamily] {
      Compute Kernel};
  \end{tikzpicture}
  \end{minipage}

  \caption{An example of the use of PyCUDA, showing the use of the
  \texttt{SourceModule} facility.
  This simple program uploads a $4\times 4$ array of single-precision
  floating point numbers, multiplies them by two on the GPU, and retrieves 
  the result.  }
  \label{lst:pycuda-demo}
\end{proglisting}

PyCUDA's interface to the `nuts and bolts' of the CUDA programming
system can be found in \texttt{pycuda.driver} and is imported here
under the alias \texttt{cuda}. The module is called \texttt{driver}
because it exposes the so-called ``driver-level'' programming
interface of CUDA, which is more flexible than the more commonly used
CUDA C ``runtime-level'' programming interface, and it has a few
features that are not present in the runtime.  (If a program uses the
triple-angle-bracket syntax for kernel invocation, it is using the
runtime interface.)

The next imported module, \texttt{pycuda.autoinit}, automatically
picks a GPU to run on, based on availability and the number, if any,
to which the \texttt{CUDA\_DEVICE} environment variable is set. It
also creates a GPU context for subsequent code to run in. Both the
chosen device and the created context are available from
\texttt{pycuda.autoinit} as importable symbols, if needed. The use of
\texttt{pycuda.autoinit} is not compulsory--if needed, users can
construct their own device-choice and context-creation methods using
the facilities available in \texttt{pycuda.driver}.

The last import in this simple example is \texttt{numpy}, the Python
array package. Since most GPU computations involve large arrays of
data, PyCUDA integrates tightly with \texttt{numpy}.

After creating a $4\times 4$ array of random single-precision floating
point numbers on the CPU in the \texttt{numpy} array identified by the
variable \texttt{a}, memory for \texttt{a} is allocated on the GPU as
\texttt{a\_gpu}. The object returned here is a
\texttt{DeviceAllocation} object, whose lifetime is coupled to that of
the GPU memory allocation, i.e. once the last reference to the
\texttt{DeviceAllocation} disappears, the object becomes eligible for
garbage collection, and once that happens, the GPU allocation also
disappears. \texttt{DeviceAllocation} objects can also be cast to
integer for pointer arithmetic. Once memory is allocated, the contents
of \texttt{a} are copied from the host to the device (hence
``\texttt{htod}''). Observe that no explicit error checking occurs at
any stage of this program. If an error is encountered, a subclass of
\texttt{pycuda.driver.Error} is raised as an exception.

Now that the data is prepared, the most interesting part of the
program begins, in which the source code for the GPU kernel is passed
to the constructor of \texttt{SourceModule}. At this point, the NVIDIA
compiler is invoked, coupled with the caching mechanism as described
in Section \ref{sec:core-method}, at the end of which the user obtains
a handle to a \texttt{driver.SourceModule} representing the binary
code uploaded to the GPU.  From this module handle, the user may then
obtain a \texttt{Kernel} handle by means of the \texttt{get\_function}
method. Observe that C++ name mangling\footnote{``Name mangling''
facilitates function overloading in C++ and represents the encoding of
signature information in the symbol name used for a function. Mangling
methods vary by compiler and operating system ABI.} would generally
make the symbol names passed to \texttt{get\_function} complicated and
non-human-readable.  For this reason, PyCUDA automatically wraps the
code passed to \texttt{SourceModule} in an \texttt{extern "C"}
declaration. If the use of C++ features is desired, the keyword
argument \texttt{no\_extern\_c} may be passed to \texttt{SourceModule}
to avoid this. In this case, any \texttt{\_\_global\_\_} entry points
should be declared \texttt{extern "C"} by hand.

The \texttt{Kernel} handle, once it has been obtained, can then be
called like any other function. Keyword arguments \texttt{grid} and
\texttt{block} determine the size of the computational grid and the
thread block size. \texttt{DeviceAllocation} instances may be passed
directly to kernels, but other arguments incur the problem that PyCUDA
knows nothing about their required type. There are two ways to address
this: First, pass all such arguments as \texttt{numpy} sized scalars,
such as \texttt{numpy.float32(5.7)} for single-precision floating
point, or \texttt{numpy.intp(p)} for pointer-sized integers.
Alternatively, one may use \emph{prepared kernel invocation}, in which
the user informs PyCUDA explicitly about the kernel's argument types.
In this case, the invocation line would need to be changed as follows:
\begin{lstlisting}
func.prepare("P", block=(4,4,1))
func.prepared_call((1, 1), a_gpu)
\end{lstlisting} 
Note that the
\texttt{prepare} method only needs to be called once.  Its first
argument represents the kernel's arguments as a type string as
accepted by the \texttt{struct} module in Python standard library,
e.g. \texttt{"P"} in the example specifies a single pointer argument.
The prepared call styles also differs in that the thread block size is
set at \texttt{prepare} time (but can be changed), and the grid
dimension may be set with each call. Prepared kernel invocation is
also slightly faster than the explicitly-sized invocation style.

In this example, the function of the GPU code itself is trivial--for a
$4\times 4$ block size and a single-element grid, each entry of the
corresponding array on the GPU is multiplied by two. As the kernel is
invoked, the kernel call enters the queue on the GPU, but, like in the
CUDA ``runtime'' interface, the invocation returns immediately and
does not wait for completion on the GPU. An argument
\texttt{stream} may be passed in either calling style to specify a
\texttt{Stream} in which execution is to take place.

At the end of this simple example, memory is allocated on the CPU and
results are transferred back, to be printed alongside the original
array. Again, like in the CUDA runtime interface, all
\texttt{memcpy\_*} functions enqueue the transfer and wait until it
completes. If this is not desired, separate \texttt{memcpy\_*\_async}
functions exist.

\subsection{What Comes in the Box}
\label{sec:pycuda-builtins}
After the introduction to the basic method of programming GPUs
with PyCUDA, this section seeks to make the reader aware of further
built-in facilities aimed at making GPU programming easier.

\subsubsection{GPU Arrays}

\begin{proglisting}[h]
  \begin{center}
\begin{lstlisting}
import pycuda.autoinit
import pycuda.gpuarray as gpuarray
import numpy

a_gpu = gpuarray.to_gpu(
    numpy.random.randn(4,4)
        .astype(numpy.float32))
a_doubled = (2*a_gpu).get()
print a_doubled
print a_gpu
\end{lstlisting}
  \end{center}

  \caption{
  An example performing the same function as Listing \ref{lst:pycuda-demo}, but
  using \texttt{GPUArray}s.
  }
  \label{lst:pycuda-demo-gpuarray}
\end{proglisting}
The example presented above, as an element-wise vector operation,
represents not only an easy introduction, but also a common use case
of GPU computations, perhaps in the form of an auxiliary step between
other calculations. For this reason, PyCUDA supplies an array object,
\texttt{pycuda.gpuarray.GPUArray} that shrinks the code of the example
of Listing \ref{lst:pycuda-demo} to that of Listing
\ref{lst:pycuda-demo-gpuarray}. Just like \texttt{numpy} arrays, but unlike
memory allocated using \texttt{DeviceAllocation}, \texttt{GPUArray}s
know about their shapes and data types. They support all arithmetic
operators and a number of methods and functions, all patterned after
the corresponding functionality in \texttt{numpy}. In addition, many
special functions are available in \texttt{pycuda.cumath}.  Arrays of
approximately uniformly distributed random numbers may be generated using
functionality in \texttt{pycuda.curandom}.

\subsubsection{Complex Numbers on the GPU}

In addition to CUDA's built-in support for real numbers, PyCUDA adds
seamless support for complex numbers. To enable this support, add the
line ``\texttt{\#include \textless pycuda-complex.hpp\textgreater}''
to your kernel and declare complex numbers as
\texttt{pycuda::complex\textless type\textgreater}, where
\texttt{type} may be, e.g., \texttt{double} or \texttt{float}.
PyCUDA's \texttt{GPUArray}s also natively support operations on
complex data types.

\subsubsection{Double Precision Textures}

Another area where PyCUDA improves on native CUDA is support for
double precision in texture fetches. This is easiest to achieve when
binding a \texttt{GPUArray} to a texture reference using the 
\texttt{GPUArray}'s \texttt{bind\_to\_texref\_ext} method, while
specifying the \texttt{allow\_double\_hack} keyword argument as true.
Within kernel code, one may then use the following pattern for texture
access:
\begin{lstlisting}
#include <pycuda-helpers.hpp>
texture<fp_tex_double, 1, cudaReadModeElementType> my_tex;
__global__ void f()
{
  double x = fp_tex1Dfetch(my_tex, threadIdx.x);
}

\end{lstlisting}
Observe in particular that the texture type was prefixed with
\texttt{fp\_tex\_} and the fetching function was prefixed with
\texttt{fp\_}.

\subsubsection{Efficient Evaluation of Element-wise Expressions}

\begin{proglisting}[h]
  \subfigure[
    An example of the use of the generic facility for
    element-wise operations.
  ]{
    \begin{minipage}[t]{0.48\textwidth}
\begin{lstlisting}[basicstyle=\footnotesize\sffamily]
import pycuda.gpuarray as gpuarray
import pycuda.autoinit
from pycuda.curandom import rand \
        as curand

a_gpu = curand((50,))
b_gpu = curand((50,))

from pycuda.elementwise import \
        ElementwiseKernel
lin_comb = ElementwiseKernel(
        "float a, float *x, "
        "float b, float *y, float *z",
        "z[i] = a*x[i] + b*y[i]",
        "linear_combination")

c_gpu = gpuarray.empty_like(a_gpu)
lin_comb(5, a_gpu, 6, b_gpu, c_gpu)
\end{lstlisting}
    \end{minipage}
    \label{lst:pycuda-elwise}
  }
  \subfigure[
    An example of the use of the generic facility for
    reductive (``folding'') operations.
  ]{
    \begin{minipage}[t]{0.48\textwidth}
\begin{lstlisting}[basicstyle=\footnotesize\sffamily]
import pycuda.gpuarray as gpuarray
import pycuda.autoinit
import numpy

a = gpuarray.arange(
        400, dtype=numpy.float32)
b = gpuarray.arange(
        400, dtype=numpy.float32)

from pycuda.reduction import \
        ReductionKernel

krnl = ReductionKernel(
        numpy.float32, 
        arguments="float *x, float *y"
        map_expr="x[i]*y[i]",
        reduce_expr="a+b", neutral="0")

my_dot_prod = krnl(a, b).get()


\end{lstlisting}
    \end{minipage}
    \label{lst:pycuda-reduction}
  }
  \caption{
    Examples of high-level primitives for working with
    \texttt{GPUArray}s.
  }
  \label{lst:pycuda-elwise-and-red}
\end{proglisting}

Evaluating deeply nested expressions on \texttt{GPUArray} instances
can be inefficient, because a new temporary is created for each
intermediate result.  Many programming languages offering
user-definable abstract vector types have this problem, for example
C++. C++ also offers a way of dealing with this particular issue in
the form of expression templates \citep{veldhuizen_1997}. While we
acknowledge that this is certainly a matter of taste, we strongly
prefer the striking simplicity of both use and implementation of
PyCUDA's facility, portrayed in Listing \ref{lst:pycuda-elwise}, over
the significant complexity of expression templates.

The functionality in the module \texttt{pycuda.elementwise} contains
tools to help generate and invoke kernels that evaluate complicated
expressions on one or several operands in a single pass. The
instrumental part of the example is the invocation of the
\texttt{ElementwiseKernel} constructor, in which the user provides
both a C-style argument list and a statement (or semicolon-separated
list of statements) to be executed for each value of \texttt{i}, which
is used as a formal index variable running across each element of
\texttt{GPUArray} instances passed to the \texttt{ElementwiseKernel}.
All these instances are required to have the same length.

\subsubsection{Map-Reduce}

In a similar spirit as the support for element-wise operations, PyCUDA
has built-in functionality for tree-based reductions on the GPU. To
make this even more useful, evaluating an element-wise expression
ahead of reduction is also supported. The facility thereby becomes a
simple implementation of the MapReduce procedure
\citep{dean_mapreduce_2008}.

Listing \ref{lst:pycuda-reduction} shows the implementation of a dot
product through an instantiation of the \texttt{ReductionKernel}
class. The constructor signature starts with the specification of the
result \texttt{dtype}, in this case \texttt{numpy.float32}. It then
proceeds as the \texttt{ElementwiseKernel} above by allowing a C
argument signature and an arbitrary C expression on whose results the
final reduction will be performed. As above, the formal variable
\texttt{i} represents the index from which the input element should be
read.

The reduction step is then specified by two further arguments, a
reduction expression of two formal arguments \texttt{a} and \texttt{b}
and an expression resulting in a neutral element with respect to the
reduction expression.\footnote{``Neutral element'' is mathematical terminology
for an element that turns a binary operator into an identity map.  For
example, zero is the neutral element of addition, and one is the
neutral element of multiplication.} Once all of this information is
specified, the resulting \texttt{ReductionKernel} may be called as
above to perform the reduction.

Observe that the end result of calling the \texttt{ReductionKernel}
instance is a \texttt{GPUArray} scalar still residing on the GPU. It
can be brought to the CPU by a call to its \texttt{get} method, or
used in-place on the GPU.

\subsubsection{Further Facilities}

In addition to elementwise operations and reductions, versions 2011.1
and newer of PyCUDA and PyOpenCL are able to assist the user with
the implementation of GPU-based parallel prefix sums (also known as
`parallel scan').  Further, they provide a
number of tools to the GPU implementer which we will mention
here, but for whose detailed description we refer to the reference
documentation.

The first of those is a key optimization for programs that allocate
and deallocate GPU memory at a rapid rate. Since CUDA's memory
allocation functions are relatively expensive operations, it
becomes expedient to retain already allocated memory in a GPU
computing process instead of freeing it. These retained blocks of
memory may then be reused once a similarly-sized block is requested
afterwards. PyCUDA supports this through the use of \emph{memory
pools}. These pools integrate with \texttt{GPUArray}s, whose
arithmetic operators are a good example of the need for repeatedly
freed and allocated memory of recurring sizes. In addition, a memory
pool implementation also exists for page-locked host memory.

Lastly, PyCUDA comes with a conjugate-gradient-based Krylov solver for large,
sparse linear systems and an implementation of sparse matrices on the GPU,
following \citep{bell_spmv_2009}. These GPU-based sparse matrices integrate
directly with sparse matrix support in the SciPy package
\citep{jones_scipy_2001} and provide computational performance similar to
that of NVIDIA's \texttt{Cusp} \citep{bell_cusp_2010} library.

\subsection{Code Generation: Benefits and Usage}
\label{sec:pycuda-code-generation}
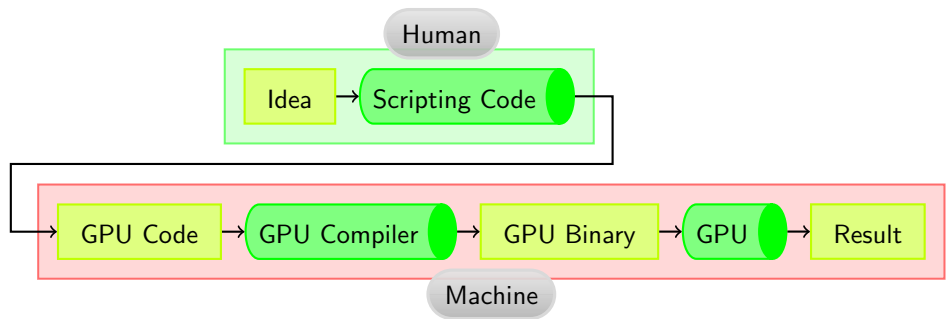
\begin{figure}
  \centering
  \beginpgfgraphicnamed{gpu-code-generation}
  \begin{tikzpicture}[
    font=\sffamily\small,
    actor/.style={cylinder,
      fill=none,draw=green,thick,
      cylinder uses custom fill=true,
      cylinder end fill=green,
      cylinder body fill=green!50,
      shape aspect=.5,
      text height=1em,
      text depth=0.5ex,
      text centered,
      },
    object/.style={
      rectangle,
      minimum height=3.5ex,
      inner xsep=3mm,
      fill=lime!50,
      draw=lime,
      text height=1em,
      text depth=0.5ex,
      thick,
      },
    hlbox/.style={
        rectangle,fill=#1!30,draw=#1,opacity=0.5,thick
    },
    button/.style={
      rounded rectangle,
      bottom color=gray!60,
      top color=gray!20,
      inner sep=2mm,
      draw=gray!30,very thick,
      },
    ]
    \node [object] (idea) at (0,0) {Idea} ;
    \node [actor,right=0.3 of idea] (py) {Scripting Code} ;
    \draw [->,thick] (idea) -- (py) ;

    \node [object] at (-2,-1.8) (cu) {GPU Code} ;
    \draw [->,thick] (py.east) -| ++(0.5,-0.9)  -| ++(-8,0) |- (cu.west) ;
    \node [actor,right=0.3 of cu] (nvcc) {GPU Compiler} ;
    \draw [->,thick] (cu) -- (nvcc) ;
    \node [object,right=0.3 of nvcc] (cubin) {GPU Binary} ;
    \draw [->,thick] (nvcc) -- (cubin) ;
    \node [actor,right=0.3 of cubin] (gpu) {GPU} ;
    \draw [->,thick] (cubin) -- (gpu) ;
    \node [object,right=0.3 of gpu] (result) {Result} ;
    \draw [->,thick] (gpu) -- (result) ;

    \begin{pgfonlayer}{background}
      \node [hlbox=green,fit=(idea) (py),inner sep=2.5mm] (human) {};
      \node at (human.north) [yshift=-0.15cm,anchor=south west,button] {Human} ;
    \end{pgfonlayer}

    \begin{pgfonlayer}{background}
      \node [hlbox=red,fit=(cu) (result),inner sep=2.5mm] (machine) {};
      \node [below=-0.15cm of machine,button] {Machine} ;
    \end{pgfonlayer}

  \end{tikzpicture}
  \endpgfgraphicnamed
  \caption{Operating principle of GPU code generation.}
  \label{fig:gpu-code-generation}
\end{figure}

As early as 30 years ago, the Lisp community observed that \emph{code
is data}, and that using code itself as the object of computation can
be greatly beneficial.  One key benefit of PyCUDA and PyOpenCL is that
they make run-time code generation (``RTCG'') almost trivial. Figure
\ref{fig:gpu-code-generation} clarifies the workflow used in RTCG.

This section is devoted to describing a number of issues that are
commonly faced when programming a GPU. In each case, we point out how
a GPU RTCG strategy can be used to address these issues.

\subsubsection{Automated Tuning}

During the creation of a GPU program, it is natural for the programmer
to come up with a number of variants of a given code, each of which
will be observed to have certain properties regarding data layout and
computation speed. The conventional approach to code tuning then calls
for the fastest variant to survive, while the others will be
discarded. This is not necessarily a desirable course of action, as
information is lost. Instead, it seems more appropriate to retain as
many of these variants as is practical, assuming that they hold at
least some promise. Further, each variant may have a number of tunable
parameters, such as loop lengths, block sizes, etc. Retaining variant
information permits choosing the best one from a reasonably sized pool
of candidates in an automated fashion, guided by some metric such as
execution speed. This is the basic premise of automated tuning, which
is trivially enabled by GPU RTCG. Further, automated tuning is not
just enabled by RTCG, it is enabled \emph{at the right time}--namely
at run time--when complete information is available. If desired, the
reader may find a few illustrative examples of the use of automated 
tuning in \citep{kloeckner_pycuda_2009}.

\subsubsection{The Cost of Flexibility}

Flexibility is commonly seen as a desirable feature of a computer
code. It should then be realized that flexibility comes at a cost:
Constants get replaced by variables, formerly fixed loop trip counts
become variable, and quite generally a compiler has less knowledge
available, making its optimizer less effective. The process of
removing this sort of flexibility by hard-coding such information into
the program, therefore, is generally frowned upon.  However, with the
availability of run-time code generation, information can be inserted
into the source of the program just in time, leading to an optimal
combination of flexibility and execution speed.

\subsubsection{High-Performance Abstractions}

Nearly all computer programs are built in `layers', where each
individual layer solves a certain subproblem and presents a more
abstract, `higher-level' interface to higher layers. This is good
engineering practice, as it allows partitioning a big problem into
many smaller ones, and it enables reuse of engineering effort.  In
some cases, such abstractions can be made uneconomical by coding
circumstance, namely when customization applies to the contents of an
inner loop. Many solutions exist to this problem, ranging from
function pointers to C++ templates, each with unique disadvantages
\citep{kloeckner_pycuda_2009}.  Once RTCG is available, this problem
also disappears as appropriate code can be generated whenever a
different requirement arises. 

\subsubsection{Generating Code}

\begin{proglisting}[h]
\begin{lstlisting}[linerange=codegen-end]
import pycuda.driver as cuda
import pycuda.autoinit
import numpy
import numpy.linalg as la
from pycuda.compiler import SourceModule

block_size = 16
thread_block_size = 32
macroblock_count = 33

a = numpy.random.randn(block_size*thread_block_size*macroblock_count)\
        .astype(numpy.float32)
b = numpy.random.randn(block_size*thread_block_size*macroblock_count)\
        .astype(numpy.float32)

a_gpu = cuda.to_device(a)
b_gpu = cuda.to_device(b)
c_gpu = cuda.mem_alloc(a.nbytes)

# codegen
from mako.template import Template

tpl = Template("""
    __global__ void add(
            ${ type_name } *tgt, 
            ${ type_name } *op1, 
            ${ type_name } *op2)
    {
      int idx = threadIdx.x + 
        ${ thread_block_size } * ${block_size}
        * blockIdx.x;

      % for i in range(block_size):
          <% offset = i*thread_block_size %>
          tgt[idx + ${ offset }] = 
            op1[idx + ${ offset }] 
            + op2[idx + ${ offset }];
      % endfor
    }""")

rendered_tpl = tpl.render(
    type_name="float", block_size=block_size,
    thread_block_size=thread_block_size)

smod = SourceModule(rendered_tpl)
# end

func = smod.get_function("add")
func(c_gpu, a_gpu, b_gpu, 
        block=(thread_block_size,1,1),
        grid=(macroblock_count,1))

c = cuda.from_device_like(c_gpu, a)

assert la.norm(c-(a+b)) == 0
\end{lstlisting}
  \caption{
    Run-Time Code Generation (RTCG) with PyCUDA using a templating
    engine.  The Example generates a piece of CUDA C from a textual
    template implementing an unrolled version of vector addition,
    using the \texttt{Mako} engine this instance.  Full context for
    the example can be found in the PyCUDA source tree as
    \texttt{examples/demo\_meta\_template.py}.
  }
  \label{lst:pycuda-meta-demo}
\end{proglisting}

We now turn to how a user might go about exploiting run-time code
generation with PyCUDA.  Since PyCUDA can natively process CUDA C
code, the objective is the generation of such code. PyCUDA makes no
assumptions about the origins of the code it processes, which allows
the logic involved in the generation to be designed to match the needs
of the application. There are, however, a few suggested ways of
generating code that we have found to cover a variety of needs.

Code generation can (and in many cases should) be seen as a text
processing task. Since one is not limited in the choice of tools with
which to perform this sort of generation, code generation typically makes use of
existing text processing tools. Generation logic itself can thus be simple and generally
responds favorably to complexity growth.

\begin{description}
  \item[Textual keyword replacement.] This simple technique performs the
  equivalent of search-and-replace on source code. It suffices for a
  surprisingly large range of use cases, such as the substitution of types and
  constants into source code at run time.  Its technological reach is increased
  by combining it with C preprocessor macros. Further contributing to its
  attractiveness, Python's standard library can perform keyword substitution
  without relying on external software.

  \item[Textual Templating.] For code generation applications where control
  flow and conditionals are required, but where all code variants are textually
  related, the use of a so-called templating engine, commonly used for the
  generation of web pages, offers a natural escalation of the capabilities of
  keyword substitution. Many templating engines (and correspondingly,
  templating languages) exist.  Listing \ref{lst:pycuda-meta-demo}
  demonstrates the use of the Mako\cite{bayer_mako_2010} engine for the
  generation of a simple, partially unrolled vector addition code.
\end{description}

In addition to these methods, the authors' \texttt{codepy} package
also supports code generation from abstract syntax trees (ASTs),
however this use is somewhat cumbersome and discouraged in all but the
most demanding cases.

\section{Evaluation}
\label{sec:eval}

%


While it is difficult to obtain quantifiable data on GPU programmer
productivity and how the use of the PyCUDA and PyOpenCL packages affects
it, one of the main measures of the success of any open-source package is
the size and vitality of the community that uses and develops it.

As such, we believe that the wide existing user base of PyCUDA and PyOpenCL
represents compelling evidence that this programming model as well as its
concrete implementations are a significant improvement in the way programmers
interact with GPUs, thereby serving as an important step toward bringing
GPU computing to the mainstream.

One central point for user collaboration is each package's wiki at
\url{http://wiki.tiker.net/PyCuda} and
\url{http://wiki.tiker.net/PyOpenCL}. In a user-editable fashion, each
functions as a central collection point for installation
instructions on a variety of operating systems, frequently asked
questions, and code examples.

In addition, a number of packages have been released by community
members that build on top of PyCUDA and PyOpenCL, including
\begin{description}
  \item[PyFFT] (by Bogdan Opanchuk) An FFT package for both PyCUDA and
  PyOpenCL. Also a nice example of cross-CUDA/OpenCL code generation.
  \url{http://pypi.python.org/pypi/pyfft}.
  \item[Scikits.CUDA] (by Lev Givon and collaborators) Offers wrappers
  of the CUBLAS, CUFFT and CULA packages for numerical computation 
  on CUDA.  \url{http://pypi.python.org/pypi/scikits.cuda}.
  \item[Atomic Hedgehog] (by Cyrus Omar) Offers a higher-level
  programming interface for PyOpenCL.
  \url{http://ahh.bitbucket.org/}.
\end{description}

For a listing of projects that use PyCUDA or PyOpenCL in
production, see \citep{kloeckner_pycuda_2009} and
\url{http://wiki.tiker.net/PyCuda/ShowCase}.

\section{Availability}

PyCUDA is available from
\url{http://mathema.tician.de/software/pycuda}, and PyOpenCL is
available from \url{http://mathema.tician.de/software/pyopencl}. Both
are distributed under the liberal MIT open-source software license. 

Full documentation is available online at
\url{http://documen.tician.de/pycuda} and
\url{http://documen.tician.de/pyopencl}, respectively, and both
packages include numerous examples and automated tests. The packages
support all platforms on which Python and CUDA and/or OpenCL are available.


%
%
%

\section{Future Directions} \ednote{This is perhaps the smallest
section with some suggestions on future directions.}

\label{sec:future}


PyCUDA and PyOpenCL, as open-source projects, thrive on user feedback,
particularly feedback regarding limitations, bugs, or missing features
that users encounter.  For example, much of PyCUDA's initial feature
set emerged in support of an effort to bring discontinuous Galerkin
methods onto the GPU, as discussed elsewhere in this volume.

As part of this continuing improvement process, we have identified a
number of core areas in which we see potential for future work.  Much
work has recently been done to make PyCUDA easier to install. Part of
this effort was the elimination of the Boost C++ library as an
explicit dependency, which was released as part of PyCUDA version 0.94
and PyOpenCL 0.92 in September of 2010. We are also working on easing
integration with existing CUDA tools such as CUBLAS, CUFFT and
Thrust\citep{hoberock_thrust_2010}.  Integration and ease of migration
between PyCUDA and PyOpenCL is another key feature that we would like
to facilitate. We are further planning to improve \texttt{GPUArray}'s
capability of dealing with slices of multi-dimensional arrays.
Finally, we are planning on improving support for automated tuning in
PyCUDA and PyOpenCL by providing search algorithms on top of
user-supplied search space descriptions and speeding up exploration of
compilation-bound searches by exploiting all available CPU cores.

We hope that this chapter has managed to encourage you, the reader, to
try what we believe is a very productive, full-featured GPU
programming environment. We look forward to your questions and
comments on our mailing lists.

\subsection*{Acknowledgments}

AK's research was partially funded by AFOSR under contract number
FA9550-07-1-0422, through the AFOSR/NSSEFF Program Award
FA9550-10-1-0180 and also under contract DEFG0288ER25053 by the
Department of Energy.  The opinions expressed are the views of the
authors. They do not necessarily reflect the official position of the
funding agencies.

\bibliographystyle{plainnat}
\bibliography{main}

\end{document}